\documentstyle{elsartwb}
\input epsf.sty

\begin{document}

\begin{frontmatter}
\title{\bf Scaling of hadron masses and widths in thermal models  
for ultra-relativistic heavy-ion collisions\thanksref{gr}} 
\thanks[gr]{Research supported in part by the Polish State Committee for
        Scientific Research, grant 2 P03B 09419 }
\author{Mariusz Michalec, Wojciech Florkowski, Wojciech Broniowski}
\address{The H. Niewodnicza\'nski Institute of Nuclear Physics, \\
        ul. Radzikowskiego 152,  PL-31342 Krak\'ow, Poland}

\begin{abstract}
By means of a simple rescaling, modifications of hadron masses and
widths are incorporated into the thermal analysis of particle ratios in
ultra-relativistic heavy-ion collisions. We find that moderate, up to
$20\%$, changes of hadron masses do not spoil the quality of the
fits, which remain as good as those obtained without modifications.
The fits with the modified masses yield, however, modified values of
the optimal temperature and baryon chemical potential.  In particular,
with decreasing masses of all hadrons (except for pseudo-Goldstone
bosons) the fitted values of the temperature and the baryon chemical
potential are lowered, with the change approximately proportional to
the scaling of masses. In addition, we find that the broadening of the
hadron widths by less than a factor of two practically does not affect
the fits.

\end{abstract}
\end{frontmatter}
\vspace{-7mm} PACS: 25.75.Dw, 21.65.+f, 14.40.-n

Recent theoretical studies \cite{BM12,Cleymans,BM3,YenGor,finland}
show that the hadronic yields and ratios in ultra-relativistic
heavy-ion collisions can be described well in the framework of simple
thermal models. The thermodynamic parameters obtained in this approach
define the so-called {\it chemical freeze-out point}, i.e., a stage
in the evolution of the hadronic system when the ``chemical
composition'' is established. Thermal-model fits to the SPS data show that the
temperature at the chemical freeze-out, $T_{chem}$, as well as the
baryon density, are large. One typically obtains $T_{chem}\sim $ 170
MeV, which is close to the expected critical value for the
deconfinement/hadronization phase transition. In this situation one
may expect that hadron properties at the chemical freeze-out are
strongly modified by the presence of the hadronic environment. Indeed,
such modifications are predicted by model calculations \cite
{BR,hatlee,hatsuda,klingl,rapp1,rapp2,rapp3}, which {\it inter alia}
helps to explain the low-mass dilepton enhancement observed in the
CERES \cite{CERES} and HELIOS
\cite{HELIOS} experiments. In this paper we incorporate possible
modifications of hadron masses and widths into thermal analysis of \
particle ratios. We generalize the results of Refs. \cite{FB1,FB2} where the
problem was studied without refitting thermodynamical parameters.

In the first part we include only the mass modifications and calculate
the particle densities from the ideal-gas expression%
\footnote{The use of Eq. (\ref{ni}) is valid when the in-medium
hadrons can be regarded as good quasi-particles. A thermodynamically
consistent approach has been constructed so far only for the lowest
multiplets of hadrons
\cite{zsch,brs}. At SPS energies, however, it is crucial to include
all hadrons with masses up to (at least) 1.8 GeV. For such a
complicated system, a thermodynamically consistent approach is not at
hand at the moment.}
\begin{equation}
n_{i}=\frac{g_{i}}{2\pi ^{2}}\int_{0}^{\infty }\frac{p^{2}\ dp}{\exp \left[
\left( E_{i}^{\ast }-\mu _{chem}^{B}B_{i}-\mu _{chem}^{S}S_{i}-\mu
_{chem}^{I}I_{i}\right) /T_{chem}\right] \pm 1},  \label{ni}
\end{equation}
where $g_{i}$ is the spin degeneracy factor of the $i$th hadron, $%
B_{i},S_{i},I_{i}$ are the baryon number, strangeness, and the third
component of isospin, and $E_{i}^{\ast }=\sqrt{p^{2}+\left( m_{i}^{\ast
}\right) ^{2}}$ is the energy. The quantities $\mu _{chem}^{B},\mu
_{chem}^{S}$ and $\mu _{chem}^{I}$ are the chemical potentials enforcing the
appropriate conservation laws. In standard thermal-model fits Eq. (\ref{ni})
is used with the vacuum masses, $m_{i}^{\ast }=m_{i}$. The in-medium masses, 
$m_{i}^{\ast }$, may depend on temperature and density in a complicated way.
In order to explore possible different behavior of in-medium masses and, at
the same time, keep simplicity, we do our calculations with the meson and
baryon masses rescaled by the two universal parameters, $x_{M}$ and $x_{B}$,
namely 
\begin{equation}
m_{M}^{\ast }=x_{M}\,m_{M},\hspace{0.5cm}m_{B}^{\ast }=x_{B}\,m_{B}.
\end{equation}
An exception from this rule are the masses of pseudo-Goldstone bosons
($\pi ,K,\eta $) which we keep constant.  

Equation (\ref{ni}) is used to calculate the ``primordial'' density of
stable hadrons and resonances at the chemical freeze-out. The final
(observed) multiplicities receive contributions from the primordial
stable hadrons, and from the secondary hadrons produced by decays of
resonances after the freeze-out. We include {\it all} light-flavor
hadrons listed in the newest review of particle physics \cite{PDG},
with a few exceptions for the cases where the properties of a listed
particle are ambiguous or not known sufficiently well. Ref. \cite{PDG}
is also used to determine the branching ratios. We neglect the
finite-size and excluded volume corrections: the former are
negligible, whereas the latter do not affect the particle ratios.%
\footnote{
The last property is due to the fact that the overwhelming majority of
hadrons is heavy and may be treated as classical particles. In this
case we may replace the Fermi-Dirac (Bose-Einstein) distribution
function in (\ref {ni}) by the Boltzmann distribution function. Thus,
for equal excluded volumes of all hadrons the excluded volume
corrections factorize and cancel out in the ratios.  We have checked that
the use of the classical statistics changes our results by a few
percent only.}

\begin{figure}[t]
\epsfysize=13.5cm
\par
\begin{center}
\mbox{\epsfbox{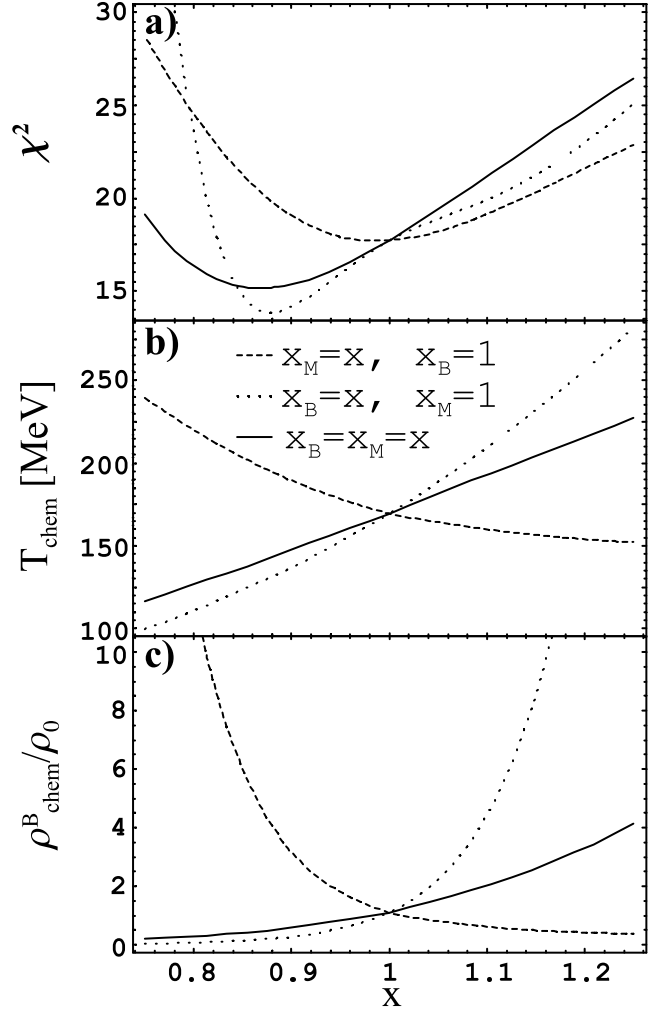}}
\end{center}
\caption{Dependence of $\protect\chi ^{2}$, and the fitted values of the
temperature and the baryon density on the scale parameter $x$. The plot
is for the Pb + Pb collisions at SPS energies, with 19 particle ratios
included in the analysis. Solid lines: all hadron masses (except for
Goldstone bosons) are scaled with $x_{M}=x_{B}=x$. Dashed lines: only meson
masses are scaled, $x_{M}=x,x_{B}=1$. Dotted lines: only baryon masses are
scaled, $x_{B}=x,x_{M}=1$. The nuclear saturation density $\protect\rho %
_{0}=0.17 \hbox{fm}^{-3}$.}
\label{fig1}
\end{figure}

For given values of $x_{M}$ and $x_{B}$ we fit the temperature, $T_{chem}$,
and the baryonic chemical potential, $\mu _{chem}^{B}$, by the minimization
the expression $\chi ^{2}=\sum_{k}\left(
R_{k}^{\,exp}-R_{k}^{\,therm}\right) ^{2}/\sigma _{k}^{2}$ , where $%
R_{k}^{\,exp}$ is the $k$th measured ratio, $\sigma _{k}$ is the
corresponding error, and $R_{k}^{\,therm}$ is the same ratio as determined
from the thermal model. The strangeness chemical potential, $\mu _{chem}^{S}$%
, and the isospin chemical potential, $\mu _{chem}^{I}$, are determined from
the requirements that: {\it i)} the initial strangeness of the system is
zero, and {\it ii)} the ratio of the baryon number to the electric
charge is the same as in the colliding nuclei.

In Fig. 1 we plot our results obtained for the experimental ratios for Pb +
Pb collisions at SPS, as compiled in Ref. \cite{BM3}. In the case $%
x_{M}=x_{B}=1$ we find: $T_{chem}=$ 169 MeV, $\mu _{chem}^{B}=$ 250 MeV, $%
\mu _{chem}^{S}=$ 65 MeV, and $\mu _{chem}^{I}=$ -9 MeV. These values are in
good agreement with those found in Ref. \cite{BM3}: $T_{chem}=$ 168 MeV, $%
\mu _{chem}^{B}=$ 266 MeV, $\mu _{chem}^{S}=$ 71 MeV, and $\mu _{chem}^{I}=$
-5 MeV. In Fig. 1 a) we give our values of $\chi ^{2}$. One can observe that
a small decrease of the meson and baryon masses, $x_{M}=x_{B}\sim 0.9$,
leads to a slightly better fit with the corresponding smaller values of the
temperature and the baryon density, as shown in Figs. 1 b) and 1 c). To
check the significance of this result we have also analyzed \ Si + Au
collisions at AGS, and S +\ Au collisions at SPS. In these two cases $\chi
^{2}$ has a flat minimum at $x_{M}\approx x_{B}\approx 1$. We thus conclude
that moderate modifications of hadron masses, say by 20\%, do not spoil the
quality of thermal fits. However, with the modified masses the thermodynamic
parameters characterizing the fits do change. For example, if we rescale
both meson and baryon masses (except for Goldstone bosons) in the same way, $%
x=x_{M}=x_{B}$, the temperature and the chemical potentials \ are to a very
good approximation also rescaled by $x$. This follows from the fact that we
study a system of equations which is invariant under rescaling of all
quantities with the dimension of energy. If we allowed also for the changes
of the masses of the Goldstone bosons, the thermodynamic parameters 
 would scale exactly as $T_{chem}(x)=x\
T_{chem}(x=0)$ and $\mu _{chem}(x)=x\ \mu _{chem}(x=0)$. In this case $\chi
^{2}$ remains constant, independently of $x$. For fixed values of the
Goldstone-boson masses, the scale invariance is broken, and $\chi ^{2}$
varies weakly with $x$, as shown in Fig. 1 a).

To account for finite in-medium widths, $\Gamma _{i}^{\ast }$, of the
resonances we generalize Eq. (\ref{ni}) to the formula \cite{BU,DMB,WW,WFN} 
\begin{eqnarray}
\hspace{-0.5cm} &&n_{i}=\int\limits_{M_{0}^{2}}^{\infty
}dM^{2}\int\limits_{0}^{\infty }dp\,\,{\frac{1}{\pi N}}{\frac{m_{i}^{\ast
}\Gamma _{i}^{\ast }}{(M^{2}-(m_{i}^{\ast })^{2})^{2}+(m_{i}^{\ast }\Gamma
_{i}^{\ast })^{2}}}  \nonumber \\
\hspace{-0.5cm} &&\times \frac{g_{i}}{2\pi ^{2}}\frac{p^{2}}{\exp \left[
\left( \sqrt{M^{2}+p^{2}}-\mu _{chem}^{B}B_{i}-\mu _{chem}^{S}S_{i}-\mu
_{chem}^{I}I_{i}\right) /T_{chem}\right] \pm 1},  \label{nigamma}
\end{eqnarray}
where $N$ is the normalization of the relativistic Breit-Wigner function, $N=%
{\frac{1}{2}}+{\frac{1}{\pi }}\arctan \left[ ((m_{i}^{\ast
})^{2}-M_{0}^{2})/(m_{i}^{\ast }\Gamma _{i}^{\ast })\right] \approx 1$. The
integral over $M^{2}$ is taken to start at the threshold $M_{0}^{2}$
corresponding to the dominant decay channel. In the limit $\Gamma _{i}^{\ast
}\rightarrow 0$ Eq. (\ref{nigamma}) obviously reduces to formula (\ref{ni}).

\begin{figure}[t]
\epsfysize=13cm
\par
\begin{center}
\mbox{\epsfbox{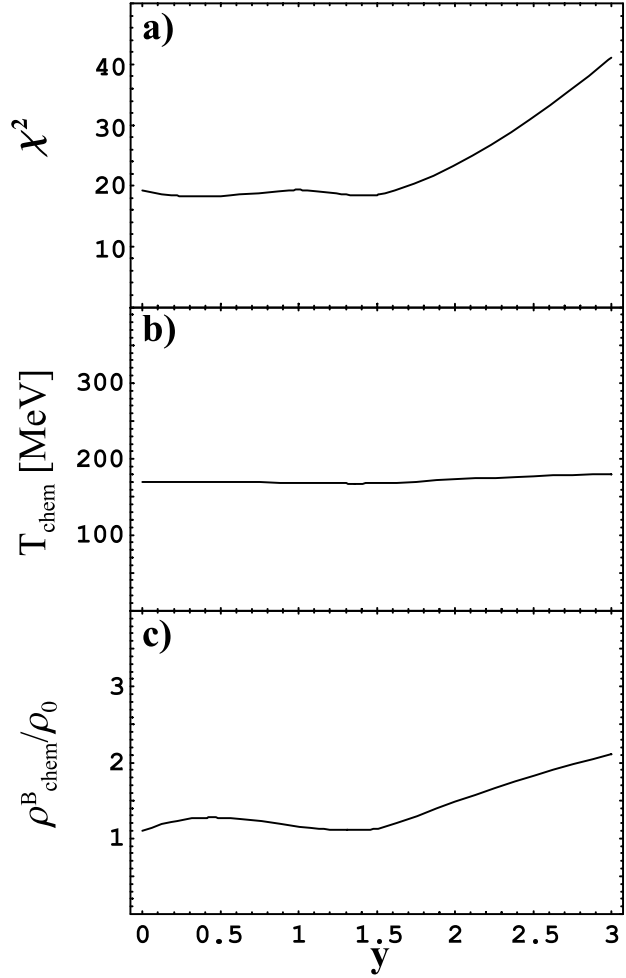}}
\end{center}
\caption{Dependence of $\protect\chi^2$, and the optimal values of the
temperature and the baryon density on the scale parameter for the widths, $y$%
.}
\label{fig2}
\end{figure}

In order to analyze the effect of broadening of hadron widths we introduce
the parameter $y$ in such a way that 
\begin{equation}
\Gamma _{i}^{\ast }=y\,\Gamma _{i}.
\end{equation}
Here $\Gamma _{i}$ are the vacuum widths, hence the case $y=1$ corresponds
to the physical widths as measured in the vacuum, and the case $y=0$
represents the situation when the widths are neglected (our previous
analysis based on Eq. (\ref{ni})). In Fig. 2 we show the results of our
fitting procedure. We observe that the inclusion of the vacuum widths does
not change the value of $\chi ^{2}$, and the values of $T_{chem}$ and $\rho
_{chem}^{B}$. An increase of the widths by a factor of 2 has also little
effect. Only for larger modifications of the widths the fit gets worse.

In conclusion we state that the thermal analysis of particle ratios allows
for scaling of hadron masses. Such changes, however, \ result in
modifications of the thermodynamical parameters for which the fits are
optimal. In particular, lowering of \ all the masses leads to a smaller
values of $T_{chem}$ and $\rho _{chem}^{B}$. This is a desired effect, since 
$T_{chem}\sim $170 MeV is large and may correspond to quark-gluon
plasma rather than to a hadron gas. Our study of the modifications of the
hadron widths shows that they have small impact on the ratios.

\end{document}